\documentclass[preprint,tightenlines,aps,showpacs,showkeys,amsmath,amssymb]{revtex4}
\usepackage{graphicx}
\begin{document} 

\title{Density dependence of symmetry free energy of hot nuclei}

\author{S. K. Samaddar$^{1}$, J. N. De$^{1}$, X. Vi\~nas$^{2}$, and
M. Centelles$^{2}$}
\affiliation{
$^1$Saha Institute of Nuclear Physics, 1/AF Bidhannagar, Kolkata
{\sl 700064}, India \\
$^2$Departament d'Estructura i Constituents de la
Mat\`eria and Institut de Ci\`encies del Cosmos,
Facultat de F\'{\i}sica, \\
Universitat de Barcelona, 
Diagonal {\sl 647}, {\sl 08028} Barcelona, Spain}


\begin{abstract}

The density and excitation energy dependence of symmetry energy and
symmetry free energy for finite nuclei are calculated microscopically in
a microcanonical framework taking into account thermal and expansion
effects. A finite-range momentum and density dependent two-body effective
interaction is employed for this purpose. The role of mass, isospin and
equation of state (EoS) on these quantities is also investigated;
our calculated results are in consonance with the available experimental
data.
\end{abstract} 



\maketitle

\section{Introduction}
 The symmetry energy is a measure of the energy involved in converting the
excess neutrons to protons in asymmetric nuclear matter. A kinetic contribution
to it comes from the associated shift of the neutron (n) and proton (p) Fermi
energies, another contribution comes from the difference between the (n-p)
interaction  and that between like pairs (n-n or p-p). Traditionally, the
symmetry energy per nucleon or the symmetry energy coefficient $C_E$ of 
infinite nuclear matter has been determined from fits of experimental 
binding energies with various versions of the liquid drop formula \cite{lun}.
But it refers only at the saturation density and at a temperature $T =0$.
Its value is usually taken to be between 30 and 35~MeV.

Understanding the details of the structure, mass and the cooling of neutron
stars \cite{fuc} or simulating the dynamics of supernovae collapse \cite{ste}
entails a knowledge of the density and temperature dependence of the symmetry energy.
The abundance of relatively heavier elements in explosive nucleosynthesis or even
the existence of exotic neutron or proton-rich nuclei produced in collisions
of radioactive nuclei have a direct lineage to this knowledge. The neutron
skin thickness of heavier nuclei has also been found to be intimately 
correlated to the density derivative of the symmetry energy 
 \cite{bro,pie,bald} as it reflects the pressure
difference on the neutrons and protons.

Collisions between nuclei at relativistic energies offer the best hope of 
studying properties related to isospin asymmetry (symmetry energy, 
symmetry free energy, etc.) of nuclear matter
at supranormal densities. Inference 
can be made there from comparison of theoretical prediction with experimental
data on symmetry energy-sensitive observables like differential flow of
neutrons and protons or from the multiplicity ratio of $\pi^-/\pi^+ $,
$K^0/K^+$, etc.\ \cite{che,fer}, but no firm conclusions can yet be made
since the experimental isospin-sensitive signals cannot be considered
very definitive \cite{lop}. At subnormal densities, studies on nuclear
multifragmentation offer a unique tool to determine the characteristics of
the nuclear symmetry energy or symmetry free energy
as a function of density and excitation energy.
In intermediate energy heavy ion collisions, a hot dilute nuclear system is
formed which expands to reach the equilibrium state and
ultimately fragments into many pieces. The produced fragments bear
signatures of the properties of the hot expanded system prior to
fragmentation. These include the excitation energy dependence of
temperature (caloric curve) and density as well as the symmetry energy
and symmetry free energy
at subnormal densities produced at different excitations. Data related
to isotopic distributions \cite{bot}, isospin diffusion \cite{ram,tsa,bali}
and isoscaling \cite{sou,she} have recently been analyzed and estimates
of symmetry coefficients at different densities and excitations   
have been obtained. These estimates give somewhat different
predictions and are also not fully conclusive.

There have been numerous studies on the symmetry energy of nuclear
matter based on the different many-body theories using various
nucleon-nucleon interactions or interaction Lagrangians \cite{che}.
These studies provide very useful tools for understanding the
properties of hot and dense nuclear matter. 
It has been noticed that the calculated
density dependence of the symmetry energy coefficient differs appreciably
depending on the choice of the theoretical models and interactions.
For the symmetry free energy of infinite nuclear matter, there 
are recent investigations done in the mean field framework \cite{xu}. 
In Refs. \cite{hor, kow}, 
the symmetry energy and symmetry entropy of very dilute nuclear matter
have been calculated exploiting virial expansion techniques where
clusterization of light fragments is taken into account. 

For finite systems, however, there are fewer available calculations 
for the symmetry energy or the symmetry free energy 
and for their dependence
on density and energy. In Ref.\ \cite{rad}, symmetry free energy coefficients
of fragments produced in nuclear multifragmentation have been calculated
from the variance of the isotopic distributions obtained in a statistical
multifragmentation model.
The present authors have performed a calculation \cite{sam} of the symmetry 
energy coefficient of finite nuclei based on the finite temperature
Thomas-Fermi (FTTF) formulation. This calculation was done microscopically
in a microcanonical framework using a finite range, momentum, and density
dependent effective interaction \cite{ban}. The calculated symmetry energy
coefficients at different excitations and densities were compared with
the available scant experimental data. 
There is an ongoing discussion regarding whether the experimental 
data for the symmetry coefficients should be connected to the symmetry 
energy or to the symmetry free energy \cite{bal}. 
We take the viewpoint that they
refer to the symmetry free energy, in accord with recent studies \cite{kow,rad} 
In the present work we calculate the symmetry free energy coefficient for a
number of nuclei in the FTTF formulation. The calculation of 
the symmetry energy coefficient in Ref.\ \cite{sam}
was done in the local density approximation (LDA). 
Calculations with some improvement over the LDA are reported
in the present paper. In addition, the dependence of the symmetry 
coefficients on the mass and isospin content of the nucleus 
as well as on the underlying EoS are considered. 

The organization of the paper is as follows. In Sec.~II, outlines of the model
used in the calculation are presented. Section III contains the results and 
discussions. Concluding remarks are given in Sec.~IV.

\section{Theoretical framework}

The methodology employed to calculate the symmetry energy and symmetry free
energy coefficients as a function of excitation energy or density is outlined
in the following. 

\subsection{Modeling the hot nucleus}

When two nuclei collide at intermediate  energy a hot nuclear system of neutrons
and protons is formed, which is  assumed to be in thermodynamic equilibrium
and can be described by a temperature $T$. The density profile of this hot
system is generated in the finite temperature Thomas-Fermi (FTTF) 
approximation with a chosen two-body effective
interaction. The details of the employed FTTF procedure are already
documented in Ref.~\cite{de} and we do not present them here.

For an expanding system pursuing the equilibrium configuration (as described
later), the surface diffuseness is likely to play an important role \cite{sob};
thus, a zero-range force like the Skyrme interaction widely used to explore
nuclear ground-state properties may not be very suitable for generating
such a density profile. It is further noted that a constrained expanded system
in the FTTF approach may lead to numerical instabilities \cite{lom,dav} and the
gradient (surface) terms in the energy density functional were replaced with
a suitable Yukawa interaction. We have therefore chosen a modified 
Seyler-Blanchard (SBM) effective interaction for the
FTTF calculations. This interaction is of finite range with momentum and
density dependence and is given by \cite{ban}
\begin{eqnarray}
v_{\rm eff}({\bf r}_1,{\bf r}_2,p,\rho)= -C_{l,u}\left [1-
\frac{p^2}{b^2}-d^2\left \{\rho ({\bf r}_1)+\rho ({\bf r}_2)\right \}^n
\right ] \frac{\exp(-r/a)}{(r/a)}.
\end{eqnarray}
An effective isospin dependence in the interaction is brought through
the different strength parameters $C_l$ for like-pair ($n-n,~p-p$) and
$C_u$ for unlike pair ($n-p$). The relative separations of the nucleons
in configuration and momentum space are given by $r=|{\bf r_1-r_2}|$
and $p=|{\bf p_1-p_2}|$.
The densities at the sites of the two interacting nucleons are given by
$\rho ({\bf r_1})$ and $\rho ({\bf r_2})$.
The parameter $a$ corresponds to the range of the interaction,
$b$ and $d$ determine its momentum and density dependence; the density
exponent $n$ controls the stiffness of the nuclear EoS.
This interaction reproduces quite well the ground-state binding
energies, root-mean square charge radii and isoscalar giant monopole
resonance energies for a host of even-even nuclei. With a density
exponent $n = 1/6$, the incompressibility of symmetric nuclear matter
$K_{\infty}$ is 238 MeV. A stiff EoS with $K_{\infty}= 380$ MeV can be
simulated with $n = 4/3$.

 In the FTTF approach, the nucleon density profile
at temperature $T$ has the form
\begin{equation}
\rho_\tau (r) =
 A_T^*(r) \, J_{1/2}\left (\eta_\tau (r)\right ),
\end{equation}
where
\begin{eqnarray}
A_T^*(r) = \frac{4\pi}{h^3}\left [2m_{\tau ,k}(r)T\right ]^{3/2},
\end{eqnarray}
and $J_K(\eta_\tau )$ is the Fermi integral
\begin{eqnarray}
J_K(\eta_\tau )=\! \!\int_0^\infty \frac{x^K}{1+\exp(x-\eta_\tau )}dx,
\end{eqnarray}
with the fugacity $\eta_\tau$ given as
\begin{eqnarray}
\eta_\tau (r)=\left [\mu_\tau - {\cal V}_\tau (r)\right ]/T.
\end{eqnarray}
In Eqs.\ (2)--(5), $\tau$ is the isospin index, $m_{\tau,k}$ the effective
$k-$mass of the nucleon coming from the momentum dependence of the
interaction, $\mu_\tau$ the chemical potentials and ${\cal V}_\tau (r)$
the effective single-particle (SP) potential (Coulomb included).

 When $\eta \ll 0$, the system is very dilute with ${\cal V} \sim 0$ and
then $\rho \sim ~e^{\mu/T}$, a constant. At large 
distances, the particle density therefore does not vanish. The pressure
at the surface is then nonzero making the system thermodynamically unstable;
the density then depends on the size
of the box in which the FTTF calculations are performed. 
This problem is overcome in the subtraction
procedure \cite{bon,sur}, where the hot nucleus, assumed to be a thermalized
system in equilibrium with a surrounding gas representing evaporated
nucleons, is separated from the embedding environment. The method is based
on the existence of two solutions to the FTTF equations, one corresponding
to the liquid phase with the surrounding gas ($lg$) and other corresponding 
to the gas ($g$) phase. The density profile of the hot nucleus in 
thermodynamic equilibrium  is given by $\rho_{\tau }= \rho_{\tau ,lg}
-\rho_{\tau ,g} $. It is independent of the box size in which calculations
are done. It also goes to zero at large distances, implying a vanishing
surface pressure. We call this the base density (it is also sometimes called
the liquid profile). The conservation of the nucleon number of each
species $N_\tau $ of the hot nucleus gives 
\begin{eqnarray}
\int\left [\rho_{\tau ,lg}(r)-\rho_{\tau ,g}(r) \right ]d{\bf r}= N_\tau .
\end {eqnarray}
The energy $E$ of the required nucleus is given by
\begin{eqnarray}
E = E_{lg}-E_g,
\end{eqnarray}
where $E_{lg}$ and $E_g$ are the total energies of the liquid-gas
system and of the gas alone.

The total entropy in the Landau quasi-particle approximation is
\begin{eqnarray}
S=-\sum_\tau \int g_\tau (\varepsilon_\tau ,T)
\left [f_\tau \ln f_\tau +(1-f_\tau )
\ln(1-f_\tau )\right ]d\varepsilon_\tau ,
\end{eqnarray}
where $f_\tau$ is the single-particle occupancy function
\begin{eqnarray}
f_\tau(\varepsilon_\tau, \mu_\tau,T)=\left [1+ \exp\{(
\varepsilon_\tau - \mu_\tau)/T\}\right ]^{-1},
\end{eqnarray}
and $g_\tau$ is the subtracted single-particle level density.
Once the energy and entropy
are known, the free energy is calculated from $F~=~E-TS$.

In the above, the description of the hot nucleus is grand canonical
obtained from the minimization of the grand potential (the temperature 
is a constant). 
In experimental conditions, however, when two nuclei collide, the hot
system is formed in isolation, its total excitation energy remains a constant.
The system might be compressed initially, resulting in a collective flow
in the decompression stage but we ignore it in the present work. The system
is microcanonical; to attain equilibrium, it expands in quest of maximum
entropy. It is, however, still possible to describe the system statistically
by an effective temperature $T$. It has the operational advantage that
it helps in defining an occupation function that can be employed in
evaluating various observables like energy, entropy, etc.

 The expansion of the hot nucleus is simulated through a self-similar 
scaling approximation for the density,
\begin{eqnarray}
\rho_\lambda (r)=\lambda^3\rho (\lambda r),
\end{eqnarray}
where the scaling parameter $\lambda$ is unity for the unbloated 
nucleus and decreases with expansion, lying in the range $0<\lambda \le 1$;
$\rho_\lambda (r)$ is the scaled density and $\rho (r)$ is the base
density profile generated in the subtracted FTTF framework. Besides its
simplicity, there is no {\it a priori} justification
for this choice, however, it has been shown that with
a harmonic oscillator potential, at relatively small temperatures, the
scaled density profiles and those generated self-consistently in a
constrained Thomas-Fermi \cite{lom} procedure
are equivalent \cite{sam1}.

One further needs to account properly for the effect of collectivity,
as the coupling of the single-particle motion with the collective degrees
of freedom \cite{boh} is not included in the FTTF procedure. This coupling
introduces an extra energy dependence in the nucleon effective mass
($m_{\omega }$, the $\omega-$mass) in addition to the $k-$mass. The total
effective mass $m^*$ can then be written as
\begin{eqnarray}
m^*=m \, \frac {m_k}{m} \, \frac {m_\omega}{m}.
\end{eqnarray}
The $\omega -$mass is surface-peaked and has values generally larger
\cite{has} than the nucleon mass $m$. This increase brings down the
excited states from higher energy to lower energy near the Fermi surface,
thus increasing the many-body density of states at low excitations.
The system can then accommodate comparatively more entropy at a given
excitation energy. The coupling of collectivity with the nucleonic
single-particle motion may thus have a significant role in getting
the equilibrium maximal entropy configuration. A self-consistent evaluation
of $m_{\omega }$ is very involved; for simplicity, we take the
same phenomenological form of Refs.\ \cite{pra,shl,de1} for it. An in-depth
presentation of our computational method of the expanded hot nucleus
with inclusion of collectivity can be found in Ref.~\cite{sam1} and
thus we do not dwell further on it here.

\subsection{Symmetry energy}

The symmetry energy $e_{sym}$ of nuclear matter
characterizes how the energy rises as
one moves away from equal numbers of neutrons and protons.
For asymmetric nuclear matter at density $\rho =\rho_n +\rho_p$
with asymmetry parameter $X=(\rho_n -\rho_p )/\rho $, the
symmetry energy is defined as
\begin{eqnarray}
e_{sym}(\rho ,T,X)=e(\rho ,T,X)-e(\rho ,T,X=0),
\end{eqnarray}
where $e$ is the total energy per nucleon of nuclear matter, given as
\begin{eqnarray}
e(\rho)=\left [ \frac {\hbar^2}{2m^*}
\tau (\rho )+{\cal E}_I(\rho )\right ] \frac{1}{\rho} \,.
\end{eqnarray}
In the above equation, the first and second terms within the square
brackets are the kinetic and potential energy densities for infinite
nuclear matter at a density $\rho $. 

The symmetry energy can be written as
\begin{eqnarray}
e_{sym}(\rho ,T,X)=C_E(\rho ,T)X^2+{\cal O}(X^4).
\end{eqnarray}
The terms beyond $X^2$ are negligible for values of $X$ one encounters 
in nuclei. The nuclear matter symmetry energy coefficient
$C_E$ is obtained from \cite{hor}
\begin{eqnarray}
C_{E}(\rho ,T)=\frac{1}{2}\frac{\partial^2}{\partial X^2}
e_{sym}(\rho ,T,X)|_{X=0}.
\end{eqnarray}
The  symmetry free energy coefficient $C_F$ can
similarly be defined as 
\begin{eqnarray}
C_{F}(\rho ,T)=\frac{1}{2}\frac{\partial^2}{\partial X^2}
f_{sym}(\rho ,T,X)|_{X=0},
\end{eqnarray}
where $f_{sym} (\rho ,T,X)$ is the symmetry free energy per nucleon
defined in the same manner as in Eq.(12) with $e$ replaced by $f$.

To compute the coefficients $C_{E}$ and $C_{F}$ in finite
nuclei we adopt the following prescription.
Once the neutron and proton equilibrium density profiles
of a nucleus with $N_0$ neutrons and $Z_0$ protons ($A_0=N_0+Z_0$)
at an excitation energy
$E^*$ and temperature $T$ are known, the symmetry energy coefficient 
can be calculated in the local density approximation (LDA) as~\cite{sam} 
\begin{eqnarray}
C_{E}(E^*)\left (\frac {N_0-Z_0}{A_0}\right )^2=\frac {1}{A_0}\int \rho (r)
\, C_{E}(\rho (r),T) X^2(r) d{\bf r}.
\end{eqnarray}
Here, $C_E(\rho (r),T)$ is the symmetry energy coefficient at temperature
$T$ of infinite nuclear matter at a density equal to the local density
$\rho (r)$ of the nucleus and 
$X(r)=\left ( \rho_n (r)-\rho_p (r) \right )/\rho (r)$ 
is the local isospin asymmetry. One can obtain analogously the
symmetry free energy coefficient $C_{F}(E^*)$ of a finite nucleus.

In the LDA, the particles at each point in space feel the potential
as if it were locally a constant. The neutron and proton potentials
in the configuration space are calculated at a temperature $T$
for infinite matter at a value of the local density $\rho (r)$ to
evaluate $C_E(\rho (r),T)$ or $C_F(\rho (r),T)$. In a finite
nucleus, these potentials at any point should also contain information
on the densities at nearby points, which in the extended Thomas-Fermi
(ETF) \cite{bra} method is taken into account by recasting the
kinetic energy density as a functional of not only the
local density but also its derivatives. The correction to the energy
density at a temperature $T$, up to second order in
$\hbar$ is \cite{bra,bar}
\begin{eqnarray}
{\cal E}_2(\rho )={\cal F}_2(\rho )+T\sigma_2(\rho),
\end{eqnarray}
where ${\cal F}_2(\rho)$ and $\sigma_2(\rho )$ are the corrections
to the free energy density and entropy density, respectively. They
are given as
\begin{eqnarray}
{\cal F}_2(\rho )& = & \frac {\hbar^2}{2m}\left \{\zeta (\eta )f 
\frac {(\nabla \rho )^2}{\rho }+\left [\frac {9}{4}\zeta (\eta )-
\frac {7}{48}\right ]\rho \frac {(\nabla f)^2}{f}\right. 
\nonumber \\   
& & \null \left. + \frac {1}{6}
(\rho \Delta f-f\Delta \rho )+\left [3 \zeta (\eta )-
\frac {5}{12}\right ]\nabla \rho \cdot\! \nabla f \right \},
\end{eqnarray}
and
\begin{eqnarray}
\sigma_2(\rho )=-\frac {\hbar^2}{2m} \frac {\nu (\eta )}{T} \left \{
f\frac {(\nabla \rho )^2}{\rho }+\frac {9}{4}\rho \frac {(\nabla f)^2}
{f}+3 \nabla \rho \cdot\! \nabla f\right \}.
\end{eqnarray}
In the above two equations, 
$\rho$ refers to the local density $\rho (r)$ and
$f={m} / {m^*(r)}$ is a functional of $\rho$. The quantity $\zeta (\eta )$, 
to a good approximation, is
\begin{eqnarray}
\zeta (\eta ) \simeq \frac{1}{36} \left [1+2/\sqrt{1+e^\eta}\right ],
\end{eqnarray}
and
\begin{eqnarray}
\nu (\eta )=-3\frac {J_{1/2}(\eta )}{J_{-1/2}(\eta )} \frac {d\zeta }{d\eta }.
\end{eqnarray}
The corrections ${\cal E}_2(\rho )$ and ${\cal F}_2(\rho )$ are added to the
local energy and free energy densities perturbatively, in the spirit of 
variational Wigner-Kirkwood theory \cite{cen}, to calculate the improved
symmetry energy and symmetry free energy coefficients.

\subsection{Isotopic scaling and symmetry free energy}

It has been observed by various experimental groups \cite{tsa2,xu1,she1,
tsa1,ono} that the logarithm of the ratio $R$ defined as 
\begin{eqnarray}
R~=~ Y_2(N,Z)/Y_1(N,Z),
\end{eqnarray}
where $Y_1$ and $Y_2$ are the yields of a particular fragment with
$N$ neutrons and $Z$ protons from two different fragmenting sources
differing in the neutron-proton ratio at the same temperature follow a
relation of the type
\begin{eqnarray}
\ln~R~=~K+(\alpha N +\beta Z).
\end{eqnarray}
This observation is known as isoscaling; the coefficients $\alpha $ and
$\beta $ are the parameters characterizing the isoscaling behavior and
$K$ is the normalization factor.

The parameter $\alpha $ has been related to the symmetry coefficient $C$
through the relation 
\begin{eqnarray}
\alpha =\frac{4}{T}C\left \{\left (\frac {Z_0}{A_0}\right )_2^2
-\left (\frac {Z_0}{A_0}\right )_1^2\right \},
\label{isoscal}
\end{eqnarray}
where the suffixes 1 and 2 correspond to the two fragmenting systems.
The quantities ($Z_0,A_0$)$_i$ denote the values in the fragmenting
system from whose disassembly the fragment ($Z,A$) is produced.
Various authors have derived Eq.~(\ref{isoscal})
under different approximations \cite{tsa,ono,bot} 
and the coefficient $C$ has commonly been related to
the symmetry energy coefficient $C_E$. In this interpretation, the
isospin dependence of entropy has been neglected, which may be a fair
approximation at around normal density but may not be so for low densities
as encountered in the tail region of the density profile of a nucleus at
a relatively high temperature. In some recent literature \cite{rad,kow},
the need to include the asymmetry dependence of entropy has been stressed
and the symmetry coefficient in Eq.~(\ref{isoscal}) has been interpreted as that
pertaining to the symmetry free energy. 
Furthermore, whether the symmetry coefficient refers to
the fragmenting source or to the primary fragments at freeze-out is not
fully settled.
In Ref.\ \cite{ono}, 
it is interpreted as
the symmetry coefficient of the primary fragments. In Ref.\ \cite{bot},
the basic interpretation is the same, but the properties of the fragments 
are conjectured to be modified due to ``in-medium'' effects because of
presence of other neighbouring fragments in the freeze-out 
volume. In 
the sequential
Weisskopf model in the grand canonical limit \cite{tsa} as applied
for an expanding emitting nucleus, the symmetry coefficient is linked
to that of the fragmenting source. In our present communication, we take the
symmetry coefficient to be the symmetry free energy of the 
expanded mononuclear system in its most probable configuration at a
fixed excitation energy $E^*$.

\section{Results and discussions}

\subsection{Infinite nuclear matter}

The SBM interaction, as has been mentioned earlier, reproduces well the bulk
properties of nuclei. For symmetric nuclear matter as well as for neutron
matter, the EoS obtained \cite{uma,rud} with this interaction also compares
very favorably with those calculated microscopically with realistic
interactions in a variational approach \cite{fri,wir}. In Fig.~1, we display
the symmetry energy coefficient of nuclear matter (at $T=0$) as a function of 
density. Since our focus of interest is in the subnuclear density region,
the results are presented up to $\rho \simeq 0.75\rho_0$ where
$\rho_0$ is the saturation density taken as 0.154 fm$^{-3}$,
its value for the SBM interaction.
The calculated results are seen to be well within the range
obtained in microscopic calculations \cite{fuc1} 
with different bare (Argonne v18) and effective (SLy4 and Gogny) interactions. 

Up to the saturation density the symmetry energy coefficient calculated
with the SBM interaction can be very well represented by 
\begin{eqnarray}
C_E(\rho ) \simeq  C_E(\rho_0 )\left (\frac {\rho }{\rho_0 } \right )^
\gamma ,
\end{eqnarray}
  with $C_E(\rho_0 ) =34.0$ MeV and $\gamma =0.65$.
Though the experimentally extracted value of the exponent $\gamma$ is still
fraught with some uncertainties, significant constraints on it have been 
determined from different observables in recent years.
Comparison of results from the transport
model with recent experimental data on isospin diffusion constrain the
value of $\gamma$ to around $0.69-1.05$ at subnuclear densities \cite{che}.
The neutron and proton transverse emission ratio measurements \cite{fam} 
present some new constraints on $\gamma$ somewhat larger than 0.5, whereas
measurements from isotopic distributions \cite{she} provide a value of
$\gamma$ close to 0.69. Consideration of the giant dipole resonance
properties of $^{208}$Pb  puts a constraint  
$23.3 < C_E(\rho \sim 0.1 {\rm fm}^{-3}) < 24.9$ MeV \cite{tri},
which implies a value of $\gamma \sim 0.55$.

In Fig.~2, the 
symmetry coefficients 
$C_E$ and $C_F$ as a function of density of nuclear matter are shown at 
$T=10$ MeV in the upper panel. The difference between
$C_E$ [Eq.~(15)] and $C_F$ [Eq.~(16)] is amplified with decrease in density,
in consonance with that obtained in Ref.~\cite{xu}.
This is understandable from entropy considerations.
Our calculations have been done in the mean-field model, inclusion
of cluster  formation at low densities would increase the values of
these coefficients somewhat \cite{hor}. 
We find that the equilibrium density ({\it i.e.},
the state at zero pressure) of nuclear matter 
falls off linearly with excitation energy and may be very well represented
by $\rho =\rho_0(1- 0.04 E^*/A)$, with $E^*$ expressed in MeV\@.
The symmetry coefficients $C_E$ and $C_F$ 
for different excitations at
equilibrium densities
calculated using Eqs.~(15) and (16) are shown in the lower panel of the
figure. The dependence of these coefficients with excitation is found to be
nearly linear and may be well represented as 
$C_E(E^*/A)\simeq C_E(0)(1-0.024E^*/A)$ and
$C_F(E^*/A)\simeq C_F(0)(1-0.028E^*/A)$. 
We have studied the role on $C_E$ and $C_F$ of using a soft EoS ($n$=1/6,
$K_\infty $=238 MeV) and a hard EoS ($n$=4/3, $K_\infty $ =380 MeV). The
effect of the EoS on both the coefficients in infinite nuclear
matter is found to be small except at very low densities. 

\subsection{Finite nuclei}

We have calculated the symmetry coefficients for a number of nuclei
in order to study their mass and asymmetry ($X_0~=~(N_0-Z_0)/A_0$)
dependence as a function of density and excitation energy. For the
mass dependence, we have chosen $^{197}$Au and $^{40}$S, both having
practically the same $X_0$. For the isospin dependence, we have considered
the isobar pair $^{150}$Sm and $^{150}$Cs. The relevant experimental
data on the symmetry coefficients are very few; they are available 
mostly in the mass region $A_0\sim 100-120$ \cite{she,fev}. We have 
therefore studied the nucleus $^{110}$Sn to have a comparison of 
the calculated results with the experimental data. The model is tested
further in a wider perspective; we calculate the evolution with
excitation of temperature (caloric curve) and density of this nucleus
as experimental data \cite{cib,nat} are available around this mass
number.

\subsubsection{Grand canonical approach}

In Fig.~3, the caloric curve, the central density $\rho_c$ in units
of the ground-state central density $\rho_{c,0}$ and the symmetry
coefficients $C_E$ and $C_F$ for the nucleus $^{110}$Sn are displayed
as a function of excitation energy $E^*/A$ in 
panels (a), (b), and (c), respectively. These calculations have 
been performed with the base density profile generated in the
grand canonical framework where all the excitation energy has been locked 
in the thermal mode, {\it i.e.}, there is no expansion energy.
The experimental data  for the caloric curve and densities 
correspond to medium-heavy nuclei ($100 < A_0 < 140$).
They have been taken from Ref.\ \cite{cib} for the caloric curve and 
from Ref.\ \cite{nat} for the densities.
We have also included in the figure the available experimental
data for the symmetry coefficients; the open triangles and the filled
circles are from Ref.~\cite{she} and the open and filled squares
are from Ref.~\cite{fev}. As discussed earlier, we interpret 
these data as pertaining to
the symmetry free energy coefficient. 
The data from Ref.\ \cite{she} correspond to collisions
between mass-symmetric nuclei with total mass $A_0 = 116$. The source size
was taken there to be somewhat less, $A_0 \simeq 100$, because of the  
reduction due to preequilibrium emission. The data in Ref. \cite{fev}
were extracted for collisions of $^{12}$C on $^{112,124}$Sn.
The symmetry coefficients there are given as a function of
temperature. We have expressed them as a function of excitation energy
using the Fermi-gas expression  
$E^*~=~aT^2$ with an effective level density parameter $a~=~A/10$. 
 
The present calculations are done for the fragmenting source
$^{110}$Sn to give an orientation on the excitation energy dependence
of the symmetry coefficients. As reported later, the symmetry
coefficients are found to be weakly dependent on the mass of the
fragmenting system, but somewhat sensitive to its $N_0/Z_0$ ratio. 
The dotted black line and dot-dash blue line of Fig.~3
correspond to calculations
in the LDA for $C_E$ and $C_F$, respectively. The corresponding calculations
with second order corrections incorporated are represented by the 
magenta dash line and the full red line.
The calculated caloric curve  matches very well
with the experimental data except at high excitations. The calculated
densities are, however,  overestimated. Correlation of 
the symmetry coefficients with the density is displayed in Fig.~4.
The notations used for the different lines are the same as in
the bottom panel of Fig.~3.
In panel (c) of Fig.~3, it is seen that the calculated symmetry free
energy coefficients follow the experimental trend rather well, but 
in Fig.~4 the mismatch between theory and experiment \cite{she}
becomes very apparent indicating the limitations of the
grand canonical approach.

\subsubsection{Microcanonical approach}

 The fact that the hot nuclear system formed in energetic nuclear 
collisions is an isolated system and the limitations of the grand
canonical approach exposed in Fig.~4, motivate one to describe
the evolution of the system in microcanonical thermodynamics.
In this framework, the system expands
in search of the maximal entropy configuration. In panels 
(a) and (b) of Fig.~5, the
caloric curve and the evolution of the density with excitation
energy calculated in the microcanonical approach are displayed
for the system $^{110}$Sn. The comparison of the observables with the experimental
data is now improved, showing the importance 
of the proper treatment of the expansion phase for  the
equilibrium configuration.
The bottom panel displays the symmetry coefficients $C_E$ and
$C_F$ in the LDA and also with the inclusion of second-order
corrections. The different lines have the same meaning 
as in the bottom panel of Fig.~3.
 With increase in excitation,
the importance of the second-order corrections is found
to decrease; this is attributed
to the slower fall of the density for nuclei bloated with excitation.
At higher excitations, the system becomes more expanded and dilute and
a possible enhancement of the symmetry coefficients with respect to
the present calculation may come from clustering at the surface \cite{kow,hor}.

The correlation of the symmetry coefficients
with density is displayed in Fig.~6 for the same system $^{110}$Sn. 
The notations used for the calculated results are the same as in Fig.~4.  
Allowing for the uncertainties in the experimental extraction of the 
density and of the symmetry coefficients, it is found that the calculated 
correlation follows the experimental trend well.  A noticeable improvement 
of the results over those depicted in Fig.~4 is observed.

 The dependence of the symmetry coefficients $C_E$ and $C_F$ for
finite nuclei on the EoS of the underlying nuclear interaction 
is displayed in Fig.~7 at different excitations.
We have chosen $^{110}$Sn as the representative system. 
All the calculations presented in this figure and  in Fig.~8
are done with the inclusion of the second
order corrections.  At the same
excitation, both $C_E$ and $C_F$ are larger for  the stiffer EoS\@. This
is understood from the fact that at the same excitation, the
equilibrium configuration is more compact for the  stiffer EoS
\cite{sam1}. As a whole, the symmetry coefficients are found to be
not too sensitive to the choice of the EoS we have made. 

The excitation energy dependence of the symmetry free energy coefficient $C_F$
for all the five nuclei studied is displayed in  panel (a) of Fig.~8.
The lines from top to bottom correspond to the 
systems $^{110}$Sn,  $^{197}$Au,
$^{150}$Sm, $^{150}$Cs and $^{40}$S, respectively. 
The mass and asymmetry dependence of the symmetry coefficient can
be easily inferred from the figure.
The comparison of the results for the systems $^{197}$Au and $^{40}$S
(having practically the same asymmetry) indicates the lowering
of the symmetry coefficients with decreasing mass. The lighter nucleus
has a lesser value of $C_F$ because of the predominance of
the surface effects. Similarly, the isospin or asymmetry dependence
can be inferred from the comparison of results of the isobar pair 
$^{150}$Sm and $^{150}$Cs. The lower values of the symmetry coefficient $C_F$
for the more asymmetric nucleus $^{150}$Cs can be traced down to the fact
that isobars with higher asymmetry have effectively softer 
EoS \cite{sam1}. It is seen that the results for the symmetry
coefficients for the pair $^{197}$Au and $^{150}$Sm are practically
indistinguishable. This reflects an interplay of the 
effects due to mass and
asymmetry. This is further amplified in the larger values of $C_F$
for $^{110}$Sn, which has an appreciably smaller mass than $^{197}$Au
but has also a very small asymmetry $X_0$~=~0.09.
In panel (b) of Fig.~8, the symmetry free energy
coefficients of all the  nuclei studied are displayed as function of their
equilibrium densities corresponding to different excitations. The variations
of the density correlation of the symmetry coefficients with mass and
isospin for the nuclei studied are very similar to those seen for the
excitation energy in the upper panel of the figure. The results for the
symmetry energy coefficient $C_E$ exhibit nearly the same trends with
excitation and density as $C_F$ in the present figure, and therefore we do
not display them.

As seen in Fig.~8, the excitation energy dependence of $C_F$ for all the 
five  nuclei discussed is almost linear and the results corresponding 
to each nucleus run nearly parallel. As in the case of nuclear matter, 
this dependence can be well approximated by a linear relation
$C_F(E^*/A)=C_F(0)(1- \alpha_F E^*/A)$ with
$\alpha_F \simeq 0.054$ MeV$^{-1}$. 
The same holds for  the symmetry energy coefficient $C_E$ (not shown
in the figure), for which we find
$C_E(E^*/A)=C_E(0)(1-\alpha_E E^*/A)$ 
with $\alpha_E \simeq 0.064$ MeV$^{-1}$.
The faster fall-off of the symmetry  
coefficients  
of finite nuclei with increasing excitation  
compared to those of nuclear matter 
is attributed to the comparatively lower equilibrium
density of the isolated nuclei
at the same excitation.

The density dependence of the symmetry
free energy coefficient of the  finite nuclei can  be fitted with a general 
expression of the form
\begin{eqnarray}
%
C_F(\rho )~=~\frac{\kappa_v \, (\rho/\rho_0)^{\gamma_1}}
{1+ \kappa_s \,(\rho/\rho_0)^{\gamma_2} \, A^{-1/3}}
\; (1-\kappa_{sym}~X_0^2)~,
\label{fit}
\end{eqnarray}
where $\kappa_v$ and $\kappa_s$ are the volume and surface constants 
contributing to the symmetry coefficient. The exponents $\gamma_1$
and $\gamma_2$ depict the density dependence of the volume and surface 
contributions, respectively. We have included a term $\kappa_{sym}\,X_0^2$,
with $X_0$ being the asymmetry parameter of the finite nucleus. This is done
in order to test the size of an eventual 
departure of the symmetry energy in finite systems
from the quadratic dependence on 
the asymmetry parameter which is assumed in 
the definition of the symmetry coefficients.

A least-squares fit of Eq.\ (\ref{fit}) to the calculated values
for the symmetry free energy coefficient fixing the values of
$\kappa_v$ (34 MeV) and $\gamma_1$ (0.65) to those of infinite
nuclear matter, and considering all the five systems studied in the
excitation energy range $1 \leq E^*/A \leq 10$ MeV, gives $\kappa_s
=1.46$, $\gamma_2 =0.17$ and $\kappa_{sym}=1.55$ with a root mean square
deviation $\simeq 6 \% $. The significantly lower value of $\gamma_2$
compared to the volume exponent $\gamma_1$ points to a weaker surface density
dependence. A free variation of all the five parameters improves the
least-squares fit very little compared to the 
variation of three parameters mentioned above. 
In infinite nuclear matter, the symmetry energy and symmetry free energy are 
known to be well represented with a quadratic term in the asymmetry parameter 
$X$, the quartic term being negligible. In finite nuclei, the existence of
surface and Coulomb effects may modify this scenario. The result
$\kappa_{sym}=1.55$ found indicates, however, that the effect is relatively
small for the typical values of $X_0$ in nuclei.

\section {Concluding remarks}

We have investigated the energy and density
dependence of the symmetry energy and symmetry free energy coefficients
of finite and infinite nuclear systems. The dependence of these
coefficients on the EoS, mass and isospin content of nuclei have
further been explored. The calculations are done in a microscopic
microcanonical framework using a momentum and density dependent
finite range effective interaction. The density dependence of the
symmetry energy coefficient of infinite nuclear matter calculated
with this interaction compares very well with those obtained from
other microscopic calculations.

Our main focus in the present work is to explore the density and energy
dependence of the symmetry coefficients of finite nuclei. 
 First, we have investigated the predictions of our considered model
for these coefficients for infinite nuclear
matter in the subnuclear density range. In the
density range
$0.1 < \rho /\rho_0  <1$, the symmetry energy coefficient
for nuclear matter at $T=0$ is found to be well reproduced by
$C_E (\rho ) \simeq C_E (\rho_0 ) \left (\rho /\rho_0 \right )^\gamma$
with $C_E (\rho_0 ) = 34.0$ MeV at the saturation density and
$\gamma \simeq 0.65$, well within the experimental range of  
values. For finite nuclei, 
the calculations have been performed in the local density approximation and 
improved by incorporating second order corrections 
in gradients of the neutron and proton densities perturbatively.
The calculated symmetry free energy coefficients are found to be
larger than the symmetry energy coefficients by $\sim$ 10$\% $ at
medium excitations because of the contribution from the symmetry entropy.
At low excitations, as expected, there is little difference in
the values of the two coefficients. At the highest excitation
(10 MeV/A) that we explore, the difference is $\sim 15 \% $.
The calculated coefficients $C_F$ compare favorably with the
available experimental data.

Both for infinite nuclear matter and for finite nuclei, the 
symmetry coefficients vary linearly with excitation energy;
however, for finite systems, the dependence is much stronger.
The  dependence on the EoS of the symmetry coefficients for 
nuclear matter is found to be rather weak. 
For finite systems the dependence is more noticeable and the 
symmetry coefficients decrease with the softness of the EoS.  
They are, however, not too sensitive to the EoS chosen.
The coefficients $C_E$ and $C_F$  of finite nuclei are
system dependent, an interplay of the role of mass and isospin 
is quite evident there. For the same asymmetry, the coefficients get 
smaller with smaller mass, this dependence is weak. On the other
hand, for the same mass, the coefficients show a relatively stronger
dependence on the isospin content of the nucleus.

The characterization of the density and excitation dependence of the 
symmetry term of the nuclear interaction is instrumental for the 
understanding of a plethora of phenomena in both nuclear physics and
astrophysics. This topic currently attracts much theoretical and
experimental activity. Our present calculations have been done
in the mean-field 
framework, effects beyond mean-field like clusterization at
low densities may have perceptible effects and therefore are worth
a study. The relevant experimental data are still very scarce,
the continuing experimental effort in reactions with 
neutron-rich stable nuclei and future data from reactions with exotic 
isotopes in the radioactive ion beam facilities 
would  contribute to a better understanding of these phenomena.

\acknowledgments{S.K.S. and J.N.D. acknowledge the financial support from
CSIR and DST, Government of India, respectively. M.C. and X.V. acknowledge
financial support from Grants No.\ FIS2005-03142 from MEC (Spain) and
FEDER, and No.\ 2005SGR-00343 from Generalitat de Catalunya, as well as
from the Spanish Consolider-Ingenio 2010 Programme CPAN CSD2007-00042.}

\newpage

\centerline
{\bf Figure Captions}
\begin{itemize}
\item[Fig.\ 1] The symmetry energy coefficient $C_E$ for nuclear matter
at $T=0$ as a function of density with different interactions. The full line
refers to calculations with SBM interaction; the other results are taken
from Ref.\ \cite{fuc1}.

\item[Fig.\ 2] The symmetry coefficients $C_E$ and $C_F$ for nuclear matter 
calculated with the SBM interaction,
$(a)$ shown as a function of density at $T=10$ MeV 
and $(b)$ as a function of excitation energy.

\item[Fig.\ 3] (Color online) The equilibrium temperature
$(a)$, the equilibrium
central density $(b)$, and the symmetry coefficients $(c)$
as a function of excitation energy for $^{110}$Sn. The calculations
are performed with the base density (without self-similar expansion).
In the bottom panel, the dotted black line and the 
dash-dot blue line are the symmetry coefficients $C_E$ and $C_F$,
respectively, calculated in LDA. The dashed magenta line and the solid red
line refer to $C_E$ and $C_F$ with inclusion of second order corrections.
For the experimental data points, see the text.

\item[Fig.\ 4] (Color online) Correlation of the symmetry
coefficients with density
for the system $^{110}$Sn. The calculations refer to those with the
base density. The experimental points are taken from Ref.\ \cite{she}.
The notations for the calculated results are the same as in Fig.~3(c).

\item[Fig.\ 5] (Color online) Same as in Fig.~3, but the
calculations are done with the microcanonical equilibrium density. 

\item[Fig.\ 6] (Color online) Same as in Fig.~4, but with the
microcanonical equilibrium density.

\item[Fig.\ 7] The dependence of the symmetry coefficients on the 
underlying EoS is shown for the finite system $^{110}$Sn. 

\item[Fig.\ 8] (Color online) The symmetry coefficient $C_F$ for the five
nuclei studied is shown as a function of excitation energy in the upper
panel, and as a function of density in the lower panel. From top to bottom,
the lines correspond to $^{110}$Sn, $^{197}$Au, $^{150}$Sm, $^{150}$Cs, 
and $^{40}$S, respectively.

\end{itemize}

\newpage

\vspace*{-1cm}
\begin{figure}[t]
\includegraphics[width=0.65\textwidth,angle=270,clip=false]{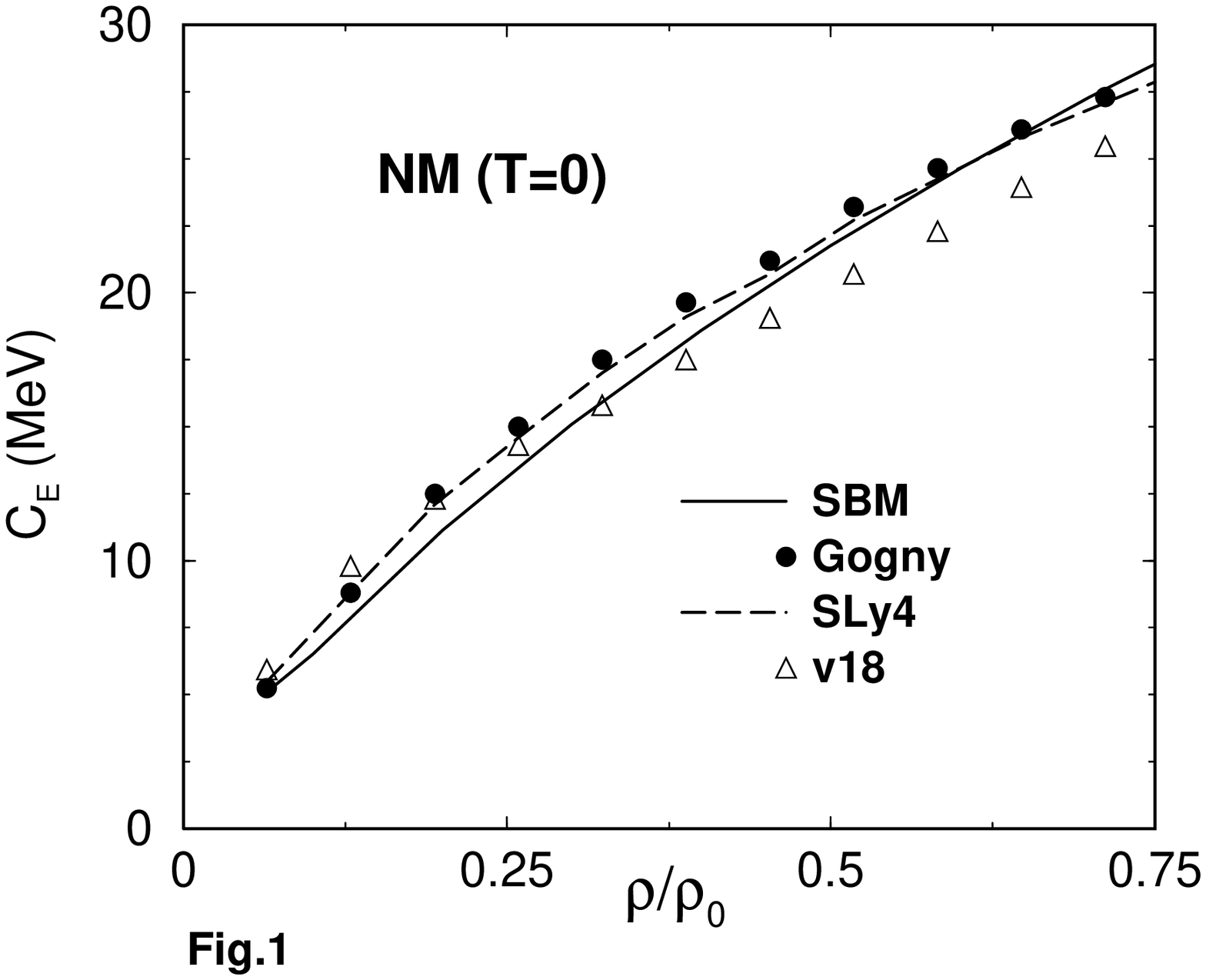}
\end{figure}

\begin{figure}[b]
\includegraphics[width=0.65\textwidth,angle=270,clip=false]{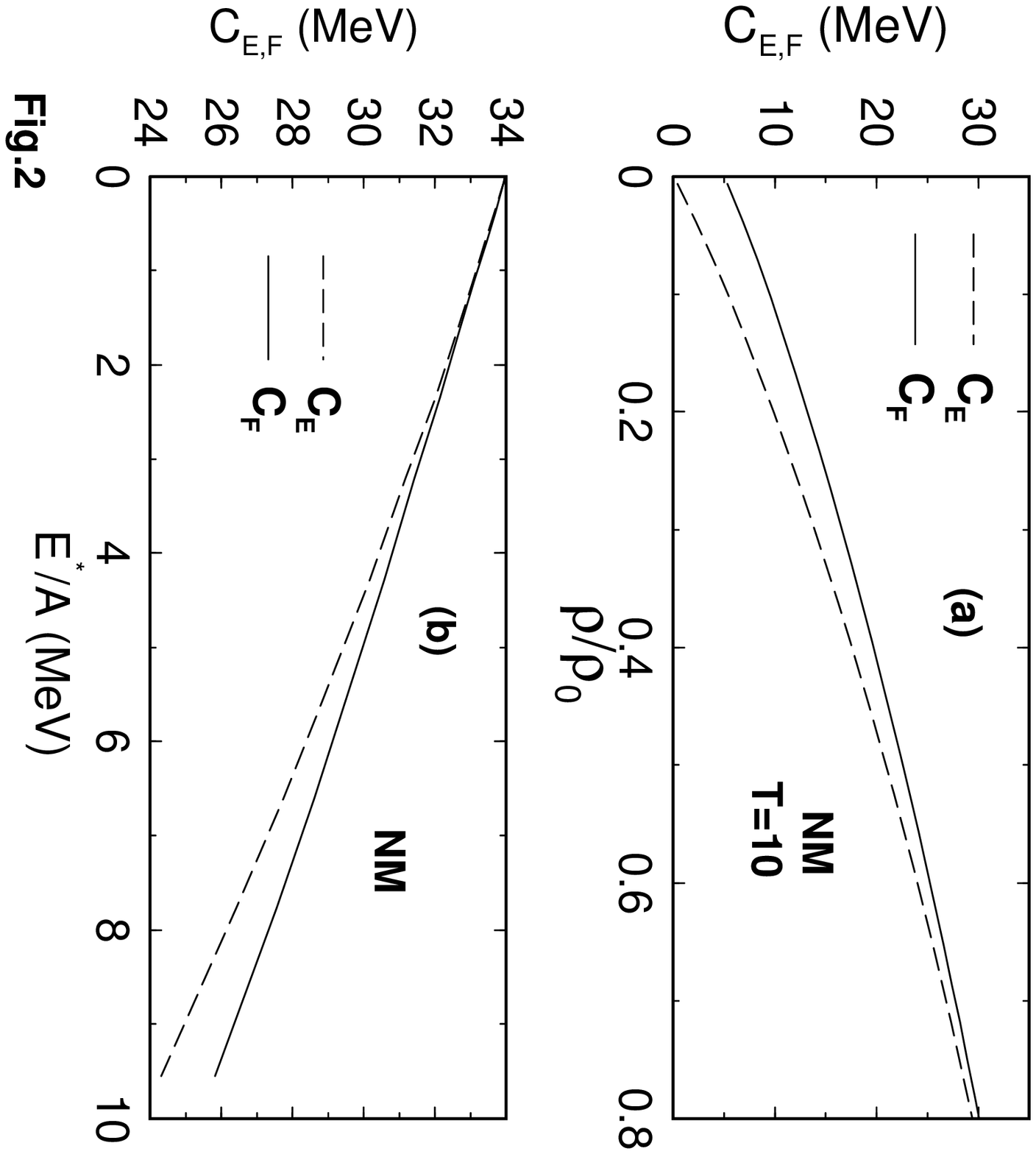}
\end{figure}

\vspace*{-1cm}
\begin{figure}[t]
\includegraphics[width=0.65\textwidth,angle=270,clip=false]{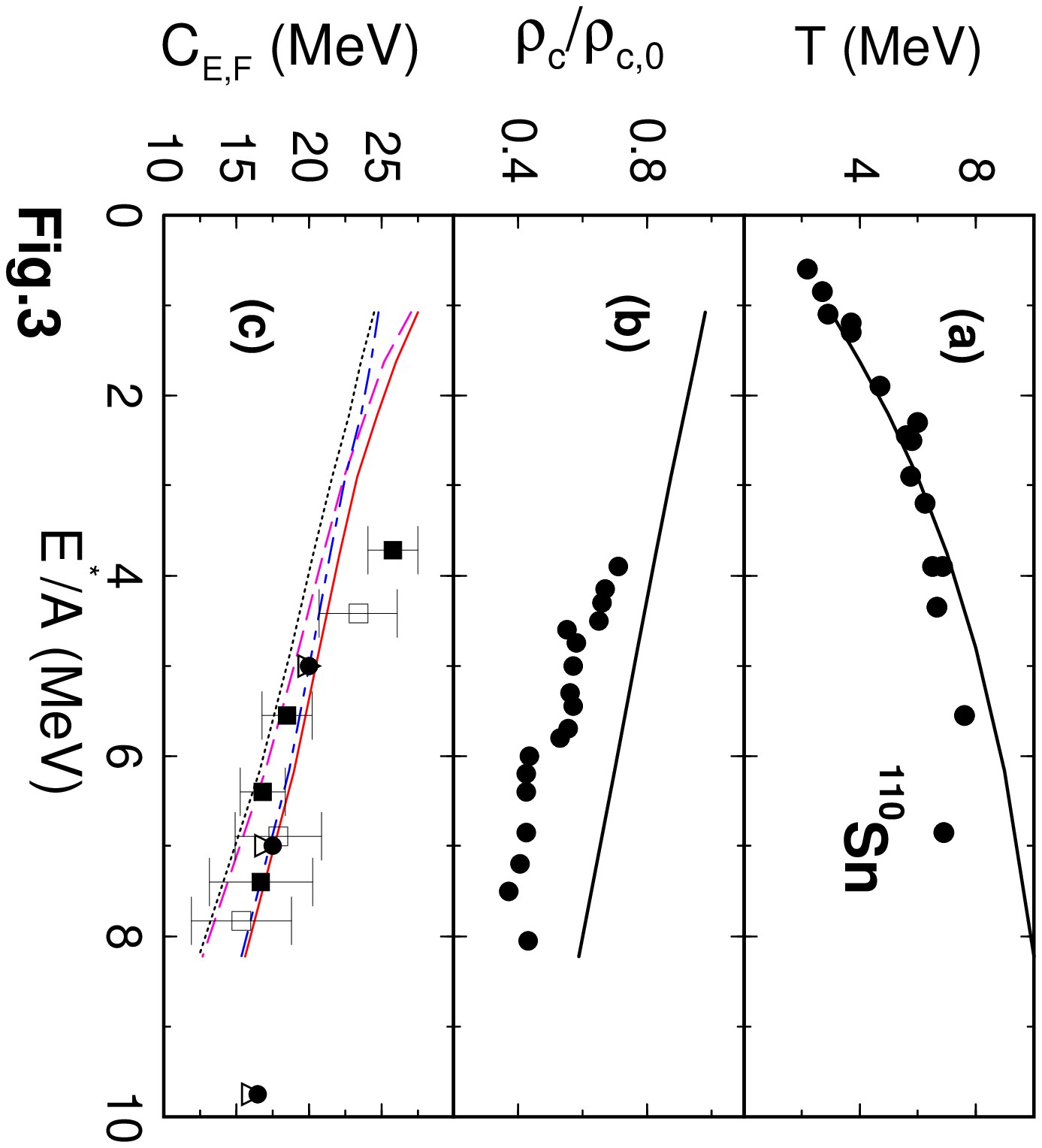}
\end{figure}

\vspace*{-1cm}
\begin{figure}[t]
\includegraphics[width=0.65\textwidth,angle=270,clip=false]{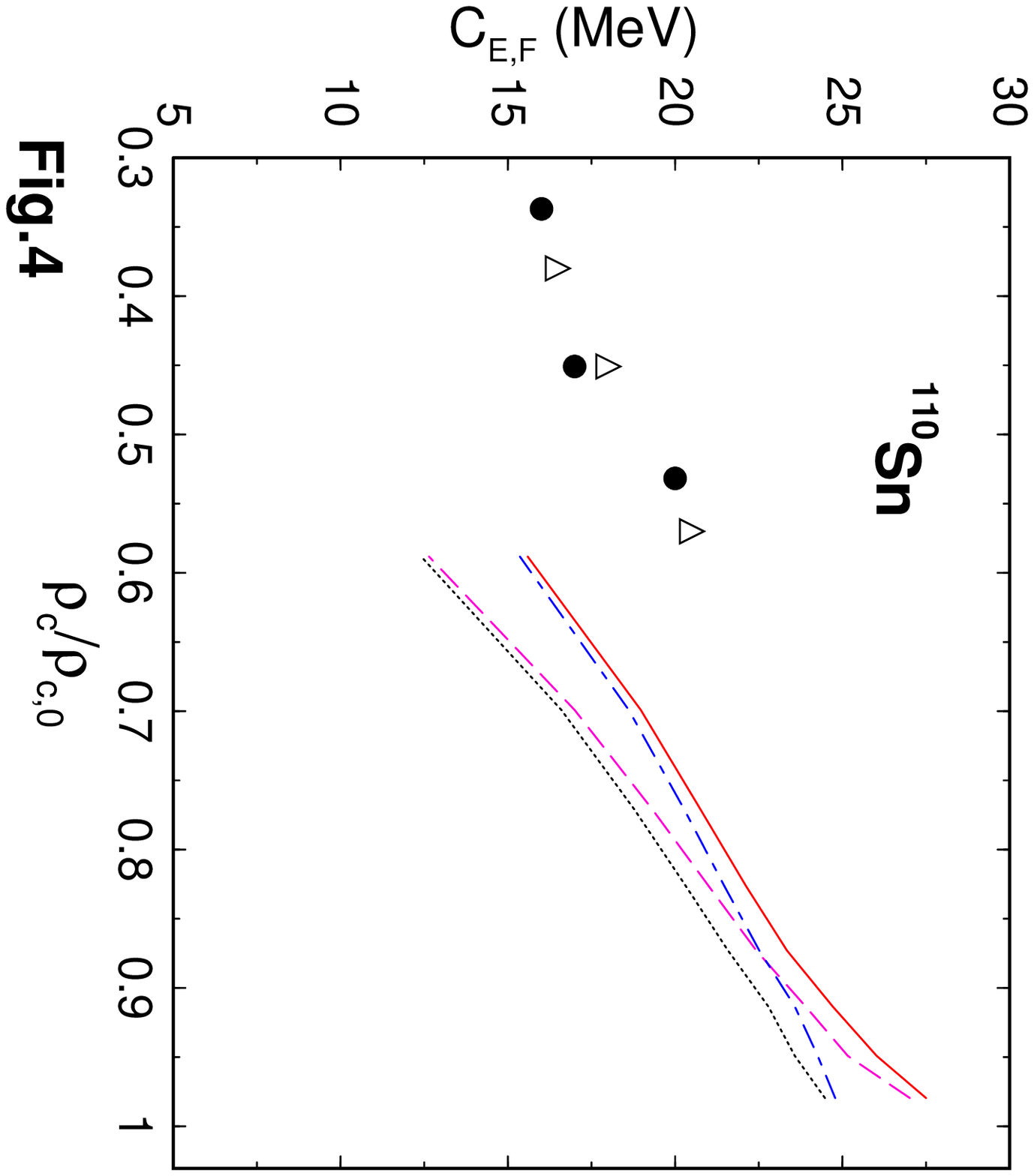}
\end{figure}

\begin{figure}[b]
\includegraphics[width=0.65\textwidth,angle=270,clip=false]{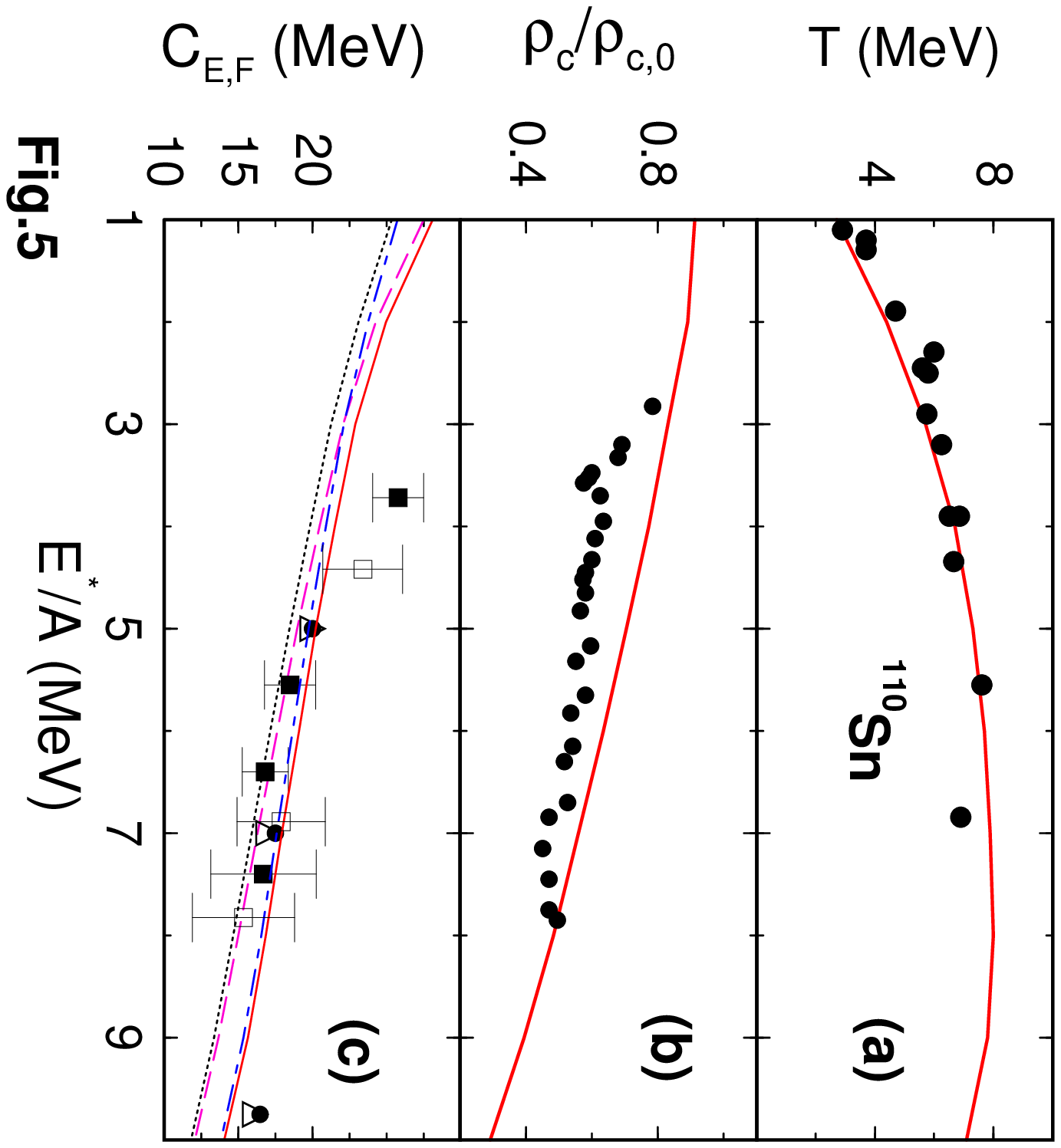}
\end{figure}

\vspace*{-1cm}
\begin{figure}[t]
\includegraphics[width=0.65\textwidth,angle=270,clip=false]{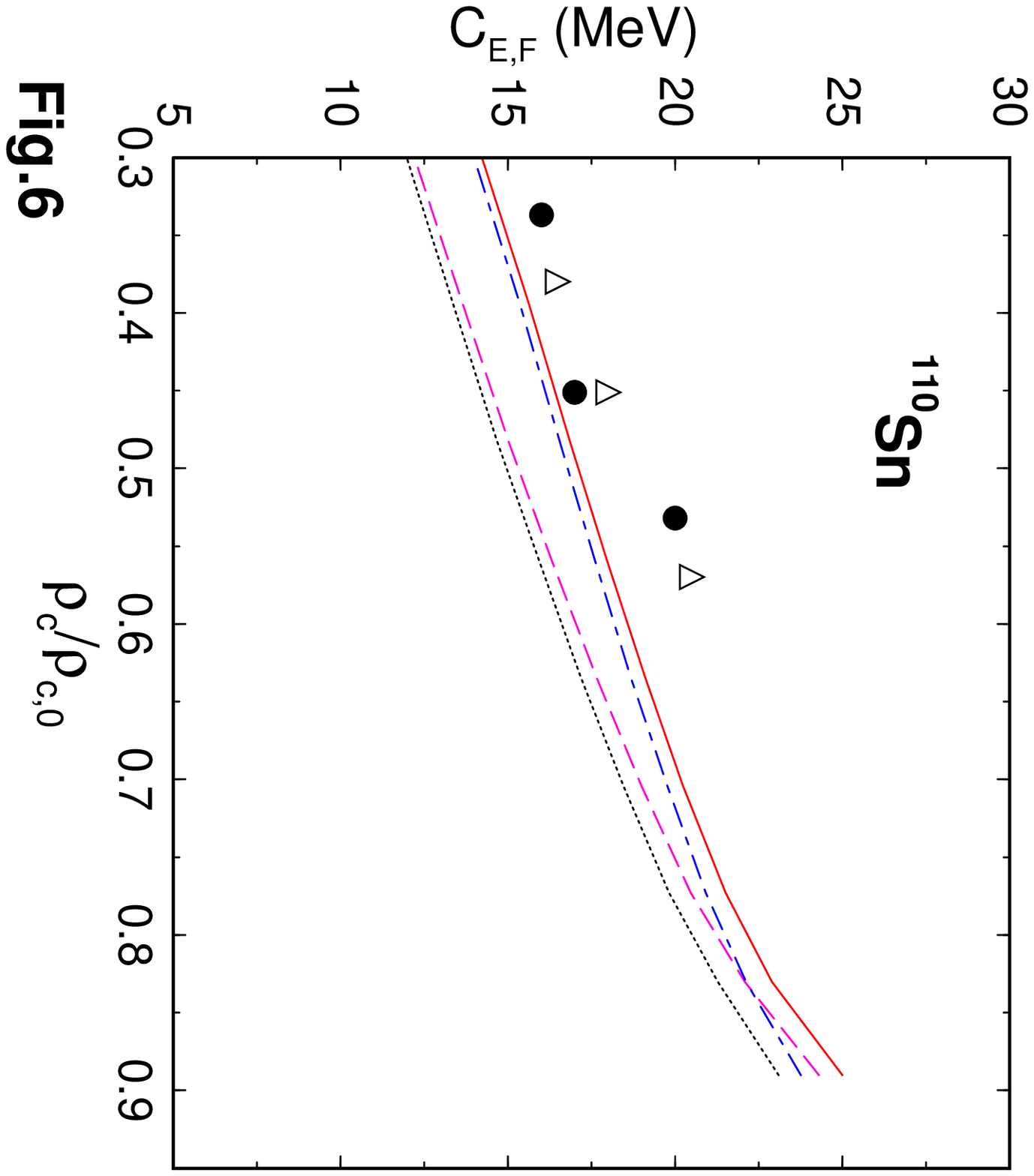}
\end{figure}

\begin{figure}[b]
\includegraphics[width=0.65\textwidth,angle=270,clip=false]{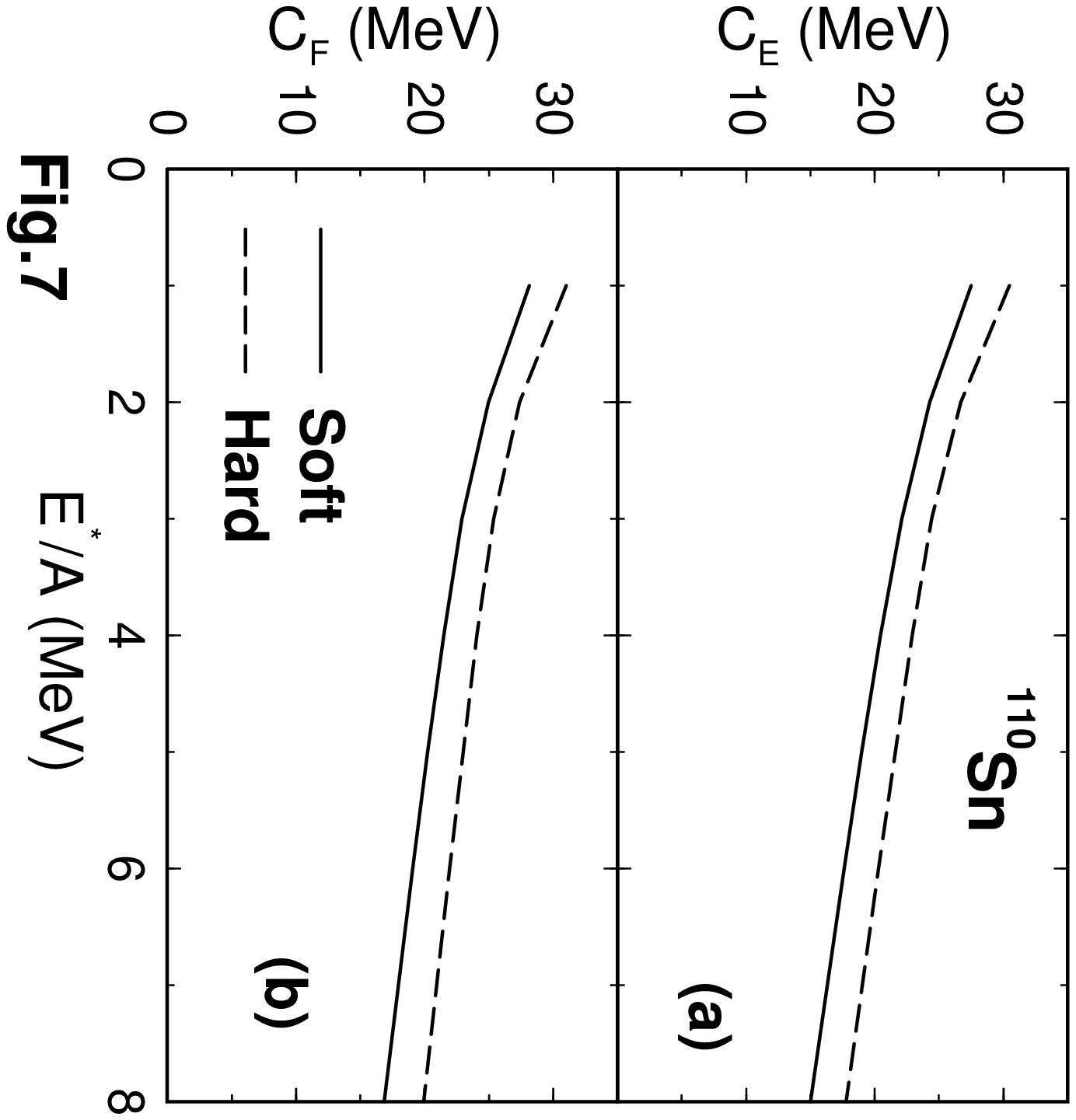}
\end{figure}

\vspace*{-1cm}
\begin{figure}[t]
\includegraphics[width=0.65\textwidth,angle=270,clip=false]{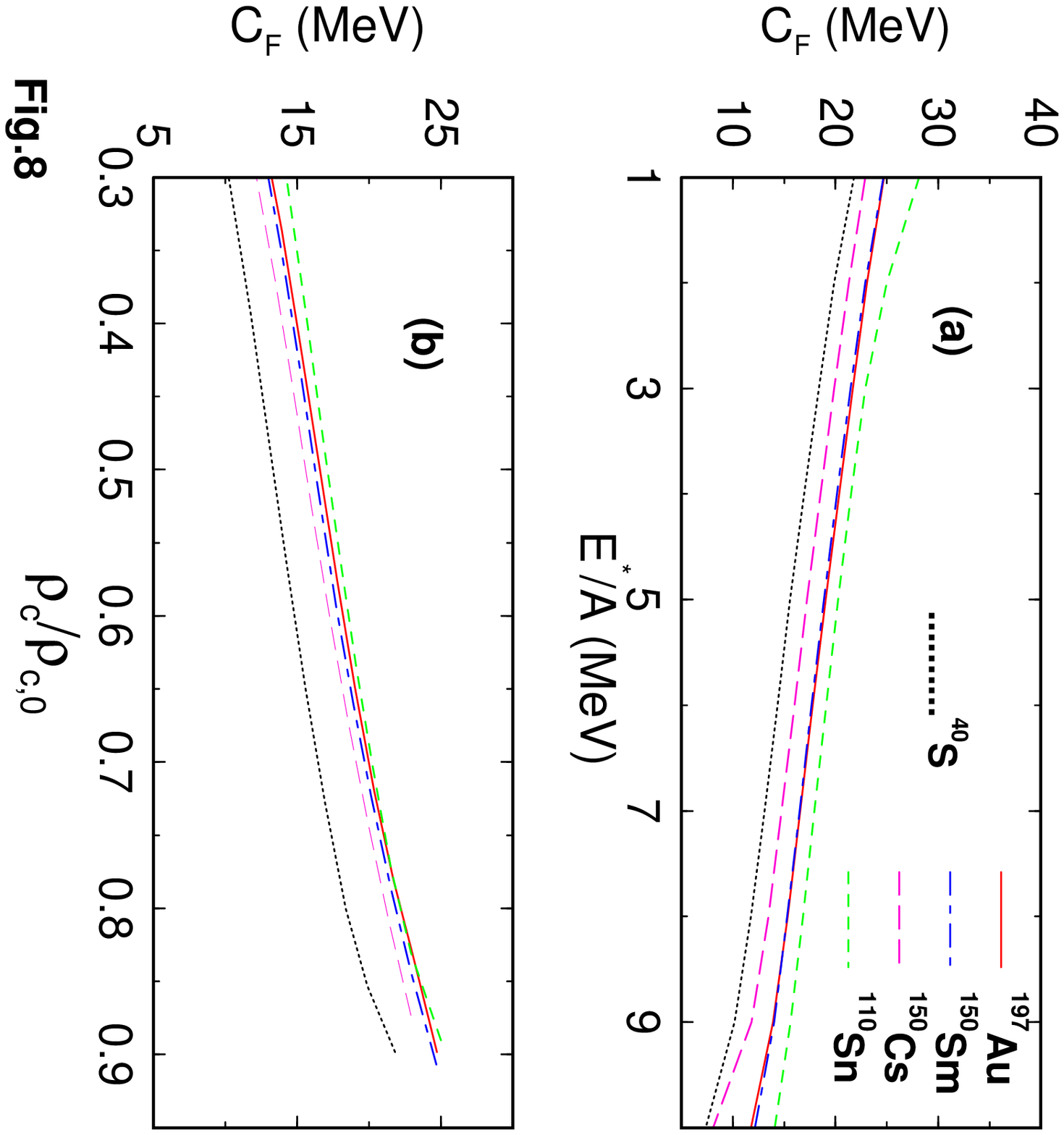}
\end{figure}

\end{document}